# Reversible Data Hiding in Encrypted Images Using MSBs Integration and Histogram Modification

Ammar Mohammadi, Mohammad Ali Akhaee and Mansor Nakhkash

*Abstract*— **This paper presents a reversible data hiding in an encrypted image that employs based notions of reversible data hiding (RDH) in a plain-image including histogram modification and prediction-error computation. In the proposed method, the original image may be encrypted by an arbitrary stream cipher. The most significant bit (MSB) of encrypted pixels are integrated to vacate room for embedding data bits. The integrated MSBs will be more robust against failure of reconstruction if they are modified for data embedding. At the recipient, we employ chess-board predictor for lossless reconstruction of the original image thanks to prediction-error analysis. Comparing to existent RDHEI algorithms, not only a separable method for data extraction is proposed, but also the content-owner may attain a perfect reconstruction of the original image without having data hider's key. Experimental results confirm that the proposed method outperforms state of the art methods.**

## I. Introduction

Reversible data hiding in plain-image (RDHPI) is drastically developed by many researchers in recent years [1]. In RDHPI, secret data is imperceptibly embedded in a plain-image in a way that the original image can be losslessly recovered after error-free secret data extraction. More presented papers in RDHPI drive from three main notions, namely difference expansion, histogram modification and lossless compression pioneered by [2], [3] and [4], respectively. Some developed studies also employ prediction-error to improve hiding capacity in a determined level of distortion. Attaining more accurate prediction, there exists sharper histogram of errors that can be modified to embed secret data. For example, schemes in [5] and [6] exploit gradient-adjusted prediction (GAP) and median edge detector (MED) predictors introduced in [7] and [8], respectively. Tsai et al. [9] present a predictor, may be denoted as local difference (LD) predictor, that computes difference between pixels intensities in a local area of the image and the most central one to bring out prediction-errors. They embed secret data via histogram modification of the prediction-errors. Sachnev et al. further present cross-dot predictor that divides an image into two "cross" and "dot" sets [10]. Dot set may be predicted using cross one and vice versa. The cross-dot predictor is also employed as a chess-board (CB) predictor in [11].

Besides RDHPI, reversible data hiding in encrypted image (RDHEI) is a solution to preserve privacy for cloud computing/storage services. In RDHEI, there are three parties: image-owner, data hider and recipient. An image-owner may not trust a channel administrator (or inferior assistant); consequently, the image-owner encrypts the image before uploading to cloud server whereas he/she is not motivated to compress his/her original-content before encryption. On the other side, data hider, i.e. the channel administrator, is not allowed to have the original-content but authorized to embed some handy data in the encrypted image. Therefore, the approach should guarantee lossless original image reconstruction and error-free data extraction at the recipient. These challenges are the cause of developing RDHEI. Schemes introduced in RDHEI may be classified into three categories namely: i) reserving room before encryption (RRBE) [12-19], ii) vacating room by encryption (VRBE) [20-26] and iii) vacating room after encryption (VRAE) [27-32].

In RRBE schemes, there exists a pre-processing before encryption that enables data hider to embed data bits in the encrypted image. Thus, in most RRBE schemes, data hider is not absolutely blind to the original content. On the other hand, most notions in VRBE are realized by encrypting some pixels intensities in a local area of the image using the same cipher byte. This approach preserves correlation between pixels employed to embed data bits. Thus, some information remains disclosed in VRBE. Except [17], the schemes presented in RRBE and VRBE are separable, which means extraction of the data bits at the recipient is not tied to decrypted information. On the other hand, in joint ones data extraction can be performed just using the decrypted marked image.

In VRAE procedure, data hider is completely blind to original information. After image encryption, data hider vacates room to embed data bits without any knowledge of the original-content. Some methods in VRAE are joint [28, 30, 31] while some others are separable [27, 29]. In [32], two different joint and separable procedures are introduced. The separable schemes in VRAE are more functional than joint VRAE schemes; in fact, they are even more functional than RRBE and VRBE schemes. Since data hider is absolutely blind to the original-content, separable VRAE methods preserve the content-owner privacy more than the others. Nevertheless, achieving high embedding capacity is more challengeable in separable VRAE than the others.

From using secret keys point of view, Chen et al. [15] classify three different schemes namely shared independent

(Corresponding author: Ammar Mohammadi.) and Mansor Nakhkash are with the Department of Electrical Engineering, Yazd University, Yazd 89195-741, Iran (e-mail: mohammadi_a@stu.yazd.ac.ir and nakhkash@yazd.ac.ir).

M. A. Akhaee is with the Department of Electrical and Computer Engineering, College of Eng., University of Tehran, Tehran, Iran, 1458889694 (e-mail: akhaee@ut.ac.ir).

secret keys (SIK), shared one key (SOK) [15] and shared no secret keys (SNK) [12, 17]. In SIK, there exist two data hider ($K_d$) and image-owner ($K_e$) keys that are shared with recipient independently while in SOK, there exists just one key. In SNK there exists no key to be shared. Most schemes in RDHEI, including the proposed one, employ SIK to manage secret keys.

Generally, similar to RDPHI, most schemes in RDHEI use the correlation of neighboring pixels in an image and further employ main notions of RDHPI such as histogram modification, difference expansion, and prediction-error computation. For example, schemes [15, 17] exploit difference expansion while Xiang and Luo employ histogram modification of the prediction-errors to reserve room before encryption [12]. Huang et al. present a new framework in RDHEI that makes it possible to use the most notions of RDH in the plain-image for the encrypted one [24]. In this self-contained scheme, any kind of prediction technique including GAP, MED, CB and LD may be used to estimate prediction-errors.

Schemes [19] and [20] use the based idea of LD predictor to embed data in the encrypted image. They realize a lossless reconstruction (LR) of the original image and error-free extraction of data bits. Qian and Zhang employ local correlation of neighboring pixels as well to reconstruct the original pixel at the recipient [29]. Using MED predictor, Yin et al. allocate some labels for almost all pixels before encryption [14]. These labels are compressed via Huffman coding and are embedded along with data bits. They improve [19] and [20] naturally thanks to the source coding algorithms. Fallahpour and Sedaaghi present a RDHPI method that apply the scheme of [3] in non-overlapped blocks of the plain-image to embed data using histogram modification of pixels in a block [33]. In the same approach, Ge et al. introduce a RDHEI method that employs histogram modification ([3]) in non-overlapped blocks of the encrypted image to vacate room for data embedding [25].

As discussed, there exist several schemes in RDHEI that employ based notions of RDHPI to embed data in the encrypted image. Here, we present a new scheme of separable VRAE that uses idea of histogram modification to vacate room after encryption.

In VRAE procedure, after a standard image encryption, data hider vacates room for data embedding without any knowledge of the original-content, i.e. data hider is completely blind to original information. At the recipient side, when the scheme of VRAE is separable, the extraction of data bits is not tied to decrypted image. Design and implementation of a separable VRAE is more challengeable but more functional than the other approaches because data hider is blind to original-content in both embedment and extraction sides. Here, we have improved other separable VRAE schemes thanks to histogram modification and MSB integration. The CB predictor is also exploited to perfect reconstruction of the original image at the recipient. Independent secret keys, $K_e$ and $K_d$ are used to encrypt the original image and data bits, respectively.

The rest of the paper is organized as follows. Related works are discussed in section II. The proposed method is presented in section III including both embedding and error-free extraction of data bits and lossless reconstruction of the original image. Section IV demonstrates the experimental results. Finally, section V concludes the paper.

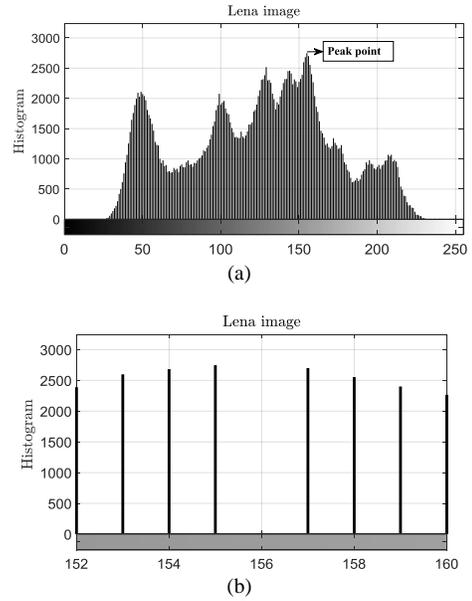

Fig. 1. (a) Histogram of the *Lena* image. (b) Vacating room to embed data bits.

## II. RELATED WORKS

For more clarification of the proposed method, we are going to explain the details of histogram modification, CB predictor and prediction-error analyzing in the following.

### A. Histogram modification

As discussed, Ni et al. [3] present a RDHPI scheme using histogram modification of the original image. In this scheme, they embed data bits in the peak point of the histogram that is the most frequent pixel in the image. Accordingly, they vacate room by histogram modification for data embedding. In this way, reversible reconstruction of the original image is possible. For instance, the peak point in the histogram of the *Lena* image (Fig. 1a) is "155". Thus, to vacate room, all intensities of the image more than "155" are added by 1 (Fig. 1b).

In our proposed procedure, we apply the idea of histogram modification to vacate room in the encrypted image (Fig. 6).

### B. Black and white prediction

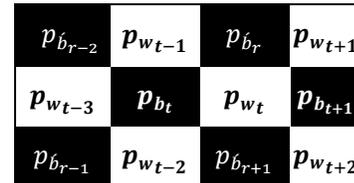

Fig. 2. Part of an image, whose pixels are divided into white and black sections.

In this section, we present two predictors related to the CB predictor [10]. As described in Fig. 2, pixels of the image can be divided to three groups, white target pixels (WTPs), black target pixels (BTPs) and black reference pixels (BRPs) that respectively denoted by $\mathbf{P}_W = \{p_{w_1}, p_{w_2}, \dots, p_{w_t}, \dots, p_{w_T}\}$, $\mathbf{P}_B = \{p_{b_1}, p_{b_2}, \ldots, p_{b_t}, \ldots, p_{b_T}\}$ and $\mathbf{P}_{\acute{B}} = \{p_{\acute{b}_1}, p_{\acute{b}_2}, \ldots, p_{\acute{b}_r}, \ldots, p_{\acute{b}_R}\}$. Data is embedded in target pixels and reference pixels remain intact during embedding process.

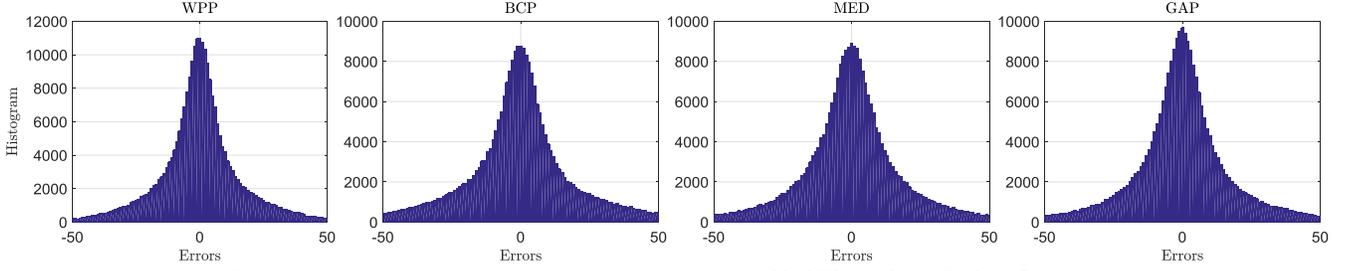

Fig. 3. Histograms of prediction-errors provided by predictors of WPP, BCP, MED and GAP for *Baboon* image.

Prediction of a white pixel, $p_{w_t}$, is done by averaging of its neighboring black ones, $\{p_{b_t}, p_{b_{t+1}}, p_{\acute{b}_r}, p_{\acute{b}_{r+1}}\}$,

$$\tilde{p}_{w_t} = \lfloor \frac{p_{b_t} + p_{b_{t+1}} + p_{\acute{b}_r} + p_{\acute{b}_{r+1}}}{4} \rceil \quad (1)$$

that denoted white plus prediction, WPP. The prediction-error, $e_{w_t}$ is computed by

$$e_{w_t} = p_{w_t} - \tilde{p}_{w_t} \quad (2)$$

In addition, a target black pixel $(p_{b_t})$ can be predicted by averaging of its neighboring black ones

$$\tilde{p}_{b_t} = \lfloor \frac{p_{\acute{b}_r} + p_{\acute{b}_{r+1}} + p_{\acute{b}_{r-1}} + p_{\acute{b}_{r-2}}}{4} \rceil \quad (3)$$

that denoted black cross prediction (BCP). In (1) and (3), $\lfloor . \rceil$ is the round function. Similarly, the prediction-error is calculated by

$$e_{b_t} = p_{b_t} - \tilde{p}_{b_t} \quad (4)$$

### C. Prediction-error analysis

According to (2) or (4), having a prediction-error and a predicted value of an original pixel, the original pixel can be definitely reconstructed.

For more explanation, let us assume a prediction-error $(e)$ that is computed from subtracting the original pixel $(\mathfrak{p})$ and its predicted value $(\tilde{\mathfrak{p}})$ as:

$$\mathfrak{p} - \tilde{\mathfrak{p}} = e \quad (5)$$

As proved in [19], even by having a range of prediction-errors, some significant bits of the original pixel will be retrieved if they substitute by data bits. In one data bit embedment, the MSB of the pixel "$\mathfrak{p}$" would be retrievable if

$$|e| < 64 \quad (6)$$

To gain a better insight, assume a pixel, $\mathfrak{p} = \mathfrak{p}_7\mathfrak{p}_6\mathfrak{p}_5\mathfrak{p}_4\mathfrak{p}_3\mathfrak{p}_2\mathfrak{p}_1\mathfrak{p}_0$, describing eight bits including the LSB, $\mathfrak{p}_0$, to the MSB, $\mathfrak{p}_7$. Having (6), $\mathfrak{p}_7$ may be replaced by a data bit in a way that it can be again retrieved at the recipient.

Employing more efficient predictor leads to sharper histogram of the prediction-errors and accordingly provides the more pixels in the image that their prediction-errors satisfy (6). Thus, there exist more MSBs of the pixels to be modified for data embedding.

In Fig. 3, the histograms of prediction-errors provided by WPP, BCP, MED and GAP predictors are demonstrated for *Baboon*. As shown, WPP provides sensibly a sharper histogram of the prediction-errors than the others. In addition, it seems GAP makes sharper histogram than MED and BCP.

If (6) is not satisfied, pixels whose prediction-errors are greater than or equal to 64 are not suitable for data embedding. Let us denote "$l$" as the number of prediction-errors which do not satisfy (6) and "$L$" as the all prediction-errors in an image. Accordingly, we define

$$f_{\text{predictor}} = p(|e| \geq 64) = l/L \quad (7)$$

as a probability of failure in reconstructing the reformed MSB

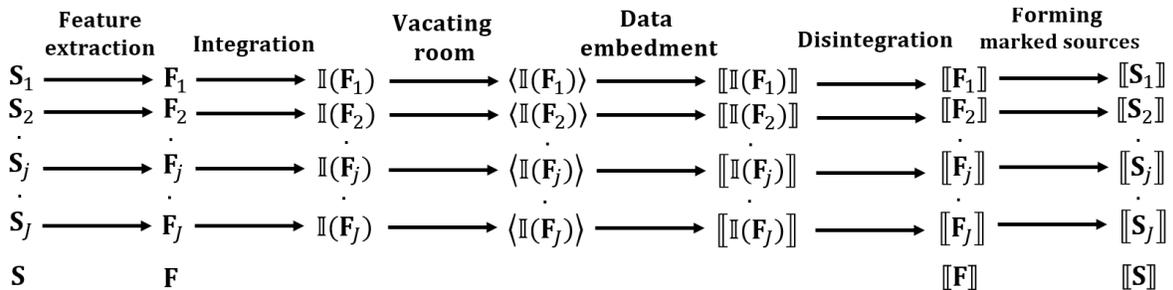

Fig. 4. Process of data embedment.

of a pixel randomly picked up. Using (7), we have $f_{WPP} = 0.0017$, $f_{BCP} = 0.021$, $f_{GAP} = 0.014$ and $f_{MED} = 0.016$ for *Baboon*. Although $f_{WPP}$ is significantly less than the others, there still exists a probability of failure in reconstructing the original pixels. As a solution, we integrate several MSBs of target pixels in order to mitigate the risk of failure. Therefore, $N_S$-MSBs, as extracted features of target pixels, are integrated using an integration model, (Fig. 5), and the integrated value is used to convey a data bit.

The amount of computed $f_{pred.}$ depends not only upon the kind of the employed predictor, but the type of a host image. The lower the entropy of an image, the lower the probability of the failure can be attained in the recovery process. Generally, smoother images have less $f_{pred.}$; for instance, *Lena* and *F16* images result in $f_{WPP} = 0$ despite images like *Baboon*.

As a consequence, we employ the best predictor, WPP, in our proposed scheme along with BCP.

## III. THE PROPOSED SCHEME

We introduce the proposed scheme and its stages including image encryption, embedding and extracting data bits, and recovering the original image.

### A. Image encryption

The encrypted image is achieved using bitwise "exclusive or" of the original one with a stream cipher provides by an encryption algorithm with an input secret key ($K_e$). Let us classify pixels forming encrypted image into three groups, namely encrypted white target pixels, $\mathbf{P}_W^e$, encrypted black target pixels, $\mathbf{P}_B^e$, and encrypted black reference pixels, $\mathbf{P}_{\tilde{B}}^e$. Encrypted BRPs remain intact during data embedding process and are used to reconstruct the black or white target pixels at the recipient.

### B. Embedding data bits

We describe embedding process in Fig. 4. Let set $\mathbf{S} = \{\mathbf{S}_1, \mathbf{S}_2, \ldots, \mathbf{S}_j, \ldots, \mathbf{S}_J\}$ be the assorted subsets of sources employed to convey data bits. The framework consist of several steps, namely, feature extraction, integration, vacating room, data embedment, disintegration and forming marked sources.

*1) Integration*

Integration model is demonstrated in Fig. 5 for a subset of $\mathbf{S}$, $\mathbf{S}_j$. Here, $\mathbf{S}$ is a set of encrypted target pixels divided into subset of $N_S$ pixels, $\mathbf{S}_j$, $j = 1,2,\ldots,J$. MSB of a pixel is taken as the feature. Therefore, $\mathbf{F}_j$ is a set of extracted features, MSBs, of corresponding pixels in $\mathbf{S}_j$. Integration is done by

$$\mathbb{I}(\mathbf{F}_j) = \sum_{k=1}^{N_S} 2^{k-1} \times F_j(k), j = 1,\ldots,J \qquad (8)$$

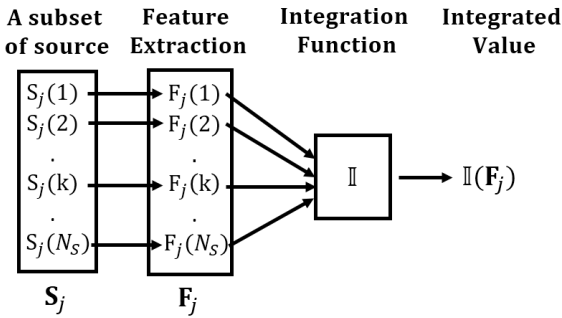

Fig. 5. Proposed model of integration.

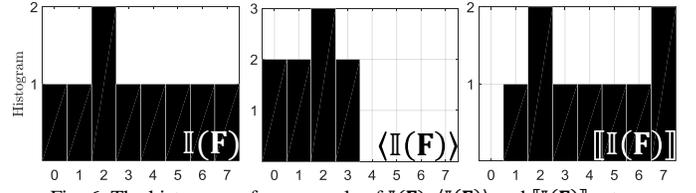

Fig. 6. The histogram of an example of $\mathbb{I}(\mathbf{F})$, $\langle\mathbb{I}(\mathbf{F})\rangle$ and $[\![\mathbb{I}(\mathbf{F})]\!]$ sets.

$N_S$ is integration parameter of a source. Employing integration, $N_S$-MSBs must be changed instead of just one MSB to embed data bit that provides more robustness against failure in the reconstruction process.

*2) Vacating room by histogram modification*

Regarding the nature of an encrypted image, the histogram of the integrated features, $\mathbb{I}(\mathbf{F})$, would have the uniform distribution. Employing the histogram modification, we vacate room in integrated features. Note that, $\mathbb{I}(\mathbf{F}_j)$ is a whole number less than $2^{N_S}$. The approach is a shrinking process that shifts each $\mathbb{I}(\mathbf{F}_j) \geq 2^{N_S-1}$ to a less one that has the most bitwise mutation than $\mathbb{I}(\mathbf{F}_j)$; in other words, $\mathbb{I}(\mathbf{F}_j)$ is replaced with its 1's complement. At the result, the shrunken amounts, $\langle\mathbb{I}(\mathbf{F})\rangle = \{\langle\mathbb{I}(\mathbf{F}_1)\rangle,\ldots,\langle\mathbb{I}(\mathbf{F}_j)\rangle,\ldots,\langle\mathbb{I}(\mathbf{F}_J)\rangle\}$, $0 \leq \langle\mathbb{I}(\mathbf{F}_j)\rangle < 2^{N_S-1}$, is achieved using Algorithm 1.

*3) Data embedment*

Having room, $J$-bits of encrypted data, $\mathbf{D}^e = \{d_1^e, d_2^e, \ldots, d_j^e, \ldots, d_J^e\}$, encrypted by $K_d$, may be embedded in $\langle\mathbb{I}(\mathbf{F})\rangle$ employing Algorithm 2 that results in marked features, $[\![\mathbb{I}(\mathbf{F})]\!]$. In this algorithm, to embed an encrypted data bit with value "1" in $\langle\mathbb{I}(\mathbf{F}_j)\rangle$, it is replaced by its 1's complement and to embed value "0" it is remained intact.

*4) Disintegration and forming marked sources*

In order to create a marked encrypted image, first, the set $[\![\mathbb{I}(\mathbf{F}_j)]\!]$ has to be disintegrated using

$$[\![F_j(k)]\!] = \mathrm{mod}\left(\left\lfloor\frac{[\![\mathbb{I}(\mathbf{F}_j)]\!]}{2^{N_S-k}}\right\rfloor, 2\right), 1 \leq j \leq J,\ 1 \leq k \leq N_S \qquad (9)$$

Substituting $[\![\mathbf{F}_j]\!]$ in the MSB of the encrypted pixels, $\mathbf{S}_j$, marked sources $[\![\mathbf{S}_j]\!], j = 1,2,\ldots,J$, are made.

Let us demonstrate the embedding process with an example. Suppose we have 27 MSBs of the encrypted pixels, $\mathbf{F} = \{0,1,1,\mathbf{0},\mathbf{0},\mathbf{0},1,0,0,\mathbf{1},\mathbf{1},\mathbf{1},0,1,0,\mathbf{0},\mathbf{1},\mathbf{0},1,1,0,\mathbf{0},\mathbf{0},\mathbf{1},1,0,1\}$, and $N_S = 3$. By integration, there exist nine whole numbers less than eight, $\mathbb{I}(\mathbf{F}) = \{3,\mathbf{0},4,7,2,\mathbf{2},6,\mathbf{1},5\}$. Employing Algorithm 1, vacating room is done by histogram modification, $\langle\mathbb{I}(\mathbf{F})\rangle = \{3,\mathbf{0},3,\mathbf{0},2,\mathbf{2},1,\mathbf{1},2\}$. Adapt values between 4 and 7 to smaller ones, room is vacated to embed $\mathcal{D}^{\mathrm{e}}$={1,**1**,0,**1**,0,**0**,1,**0**,1}. Eventually, exploiting Algorithm 2, the marked set $[\![\mathbb{I}(\mathbf{F})]\!] = \{4, \mathbf{7}, 3, \mathbf{7}, 2, \mathbf{2}, 6, \mathbf{1}, 5\}$, is attained. The histogram of $\mathbb{I}(\mathbf{F})$, $\langle\mathbb{I}(\mathbf{F})\rangle$ and $[\![\mathbb{I}(\mathbf{F})]\!]$ are depicted in Fig. 6. In disintegration, we have $[\![\mathbf{F}]\!] = $ {1,0,0, **1**, **1**, **1**, 0,1,1, **1**, **1**, **1**, 0,1,0, **0**, **1**, **0**, 1,1,0, **0**, **0**, **1**, 1,0,1}.

### C. Data extraction

At the recipient, for data extraction, we just need data hider key, $K_d$. MSBs of pixels in $[\![\mathbf{S}]\!]$ are extracted to bring out $[\![\mathbf{F}]\!]$. Considering integration model, to achieve $\mathbb{I}([\![\mathbf{F}]\!])$, subsets of $[\![\mathbf{F}]\!]$ are integrated by (8) where $\mathbf{F}_j$ is replaced by $[\![\mathbf{F}_j]\!]$. Note that, $\mathbb{I}([\![\mathbf{F}_j]\!])= [\![\mathbb{I}(\mathbf{F}_j)]\!]$.

Having $\mathbb{I}([\![\mathbf{F}_j]\!])$, extracting data bits are done employing Algorithm 3. In Algorithm 3, $j$'th bit of the encrypted data, $d_j^{\mathrm{e}}$, is extracted using $\mathbb{I}([\![\mathbf{F}_j]\!])$. Encrypted bits are decrypted using $K_d$ to bring out **D**.

### D. Recovering the original image

Reconstructing the original image may be initiated by decryption of the marked encrypted image using $K_e$. Let $[\![\mathbf{S}^{\mathrm{d}}]\!] = \{[\![\mathbf{S}_1^{\mathrm{d}}]\!], [\![\mathbf{S}_2^{\mathrm{d}}]\!], \ldots, [\![\mathbf{S}_j^{\mathrm{d}}]\!], \ldots, [\![\mathbf{S}_J^{\mathrm{d}}]\!]\}$ be the decrypted marked pixels. By replacing the MSB of the decrypted marked pixels with their 1's complement values, an alternative set, $[\![\dot{\mathbf{S}}^{\mathrm{d}}]\!] = \{[\![\dot{\mathbf{S}}_1^{\mathrm{d}}]\!], [\![\dot{\mathbf{S}}_2^{\mathrm{d}}]\!], \ldots, [\![\dot{\mathbf{S}}_j^{\mathrm{d}}]\!], \ldots, [\![\dot{\mathbf{S}}_J^{\mathrm{d}}]\!]\}$, is formed. Therefore, there exist two candidates $[\![\mathbf{S}_j^{\mathrm{d}}]\!]$ and $[\![\dot{\mathbf{S}}_j^{\mathrm{d}}]\!]$, one of them is nominated as retrieved subset, $\mathbf{S}_j^r$, that drives from a decision process is shown in Fig. 7. These two candidates are individually integrated using proposed integration model (Fig. 5) when feature set is prediction-errors. Employing neighboring pixel of a source pixel, prediction-errors may be calculated by WPP or BCP depends on position of source pixels can be white or black, respectively. For subsets of $[\![\dot{\mathbf{S}}]\!]$, prediction-errors are computed. Therefore $\mathbf{F}_j^e$ and $\dot{\mathbf{F}}_j^e$ are prediction-errors of corresponding pixels in $[\![\mathbf{S}_j^{\mathrm{d}}]\!]$ and $[\![\dot{\mathbf{S}}_j^{\mathrm{d}}]\!]$, respectively. Employing integration function that defines the summation of absolute values of the features

$$\mathbb{I}(\mathbf{F}_j) = \sum_{k=1}^{N_S} |\mathbf{F}_j(k)|, j = 1,2 \ldots, J \quad (10)$$

we compute $\mathbb{I}(\mathbf{F}_j^e)$ and $\mathbb{I}(\dot{\mathbf{F}}_j^e)$. In an alternative notation, integration may be denoted by related sources of the features, $\mathbb{I}([\![\mathbf{S}_j^{\mathrm{d}}]\!])$ and $\mathbb{I}([\![\dot{\mathbf{S}}_j^{\mathrm{d}}]\!])$, respectively. As shown in Fig. 7, for decision, the source provides less integrated prediction-errors nominated as recovered source, $\mathbf{S}_j^r$, $j = 1,2 \ldots, J$. In equal integrated prediction-errors, $[\![\mathbf{S}_j^{\mathrm{d}}]\!]$ is taken; nevertheless we probably have failure in reconstruction. If $\mathbb{I}(\mathbf{F}_j^e)$ and $\mathbb{I}(\dot{\mathbf{F}}_j^e)$ are

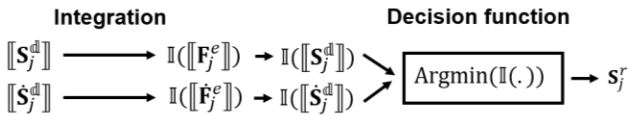

Fig. 7. Procedure of making decision to recover original set.

**Algorithm 1: Shrinking the value of $\mathbb{I}(\mathbf{F}_j)$ to $\langle\mathbb{I}(\mathbf{F}_j)\rangle$.**

  **for** $j = 1$ **to** $J$ **do**
    $\langle\mathbb{I}(\mathbf{F}_j)\rangle = \mathbb{I}(\mathbf{F}_j)$
    **if** $(\mathbb{I}(\mathbf{F}_j) \geq 2^{N_S-1})$ **then**
      $\langle\mathbb{I}(\mathbf{F}_j)\rangle = 2^{N_S} - 1 - \mathbb{I}(\mathbf{F}_j)$
    **end if**
  **end for**

**Algorithm 2: Embedding data bits.**

  **for** $j = 1$ **to** $J$ **do**
    $[\![\mathbb{I}(\mathbf{F}_j)]\!] = \langle\mathbb{I}(\mathbf{F}_j)\rangle$
    **if** $(d_j^{\mathrm{e}} == 1)$ **then**
      $[\![\mathbb{I}(\mathbf{F}_j)]\!] = |2^{N_S} - 1 - \langle\mathbb{I}(\mathbf{F}_j)\rangle|$
    **end if**
  **end for**

**Algorithm 3: Extracting data bits.**

  **for** $j = 1$ **to** $J$ **do**
    **if** $([\![\mathbb{I}(\mathbf{F}_j)]\!] \geq 2^{N_S-1})$ **then**
      $d_j^{\mathrm{e}} = 1$
    **else**
      $d_j^{\mathrm{e}} = 0$
    **end if**
  **end for**

**Algorithm 4: Risk of failure in reconstruction.**

  **if** $(\mathcal{R}_j < 16 \times N_S)$ **then**
    High risk (HiR)
  **else if** $(\mathcal{R}_j < 32 \times N_S)$ **then**
    Median risk (MeR)
  **else if** $(\mathcal{R}_j < 64 \times N_S)$ **then**
    Low risk (LoR)
  **else**
    Very low risk (VLoR)
  **end if**

close to each other, it will be a high risk in realizing LR. Therefore, the risk of failure in reconstruction can be analyzed by computing the difference between $\mathbb{I}(\mathbf{F}_j^e)$ and $\mathbb{I}(\dot{\mathbf{F}}_j^e)$.

$$\mathcal{R}_j = |\mathbb{I}(\mathbf{F}_j^e) - \mathbb{I}(\dot{\mathbf{F}}_j^e)| \quad (11)$$

The larger $\mathcal{R}_j$, the lower risk would be obtained and vice versa. Employing $\mathcal{R}_j$ and $N_S$ in Algorithm 4, we can define four classes of risk namely: high risk (HiR), median risk (MeR), low risk (LoR) and very low risk (VLoR). Therefore, in each subset of $N_S$-pixels, we have a risk analysis. The bigger $N_S$ is chosen, the more accurate evaluation of the risk would be realized.

### E. Overall view of the proposed method

In Fig. 8a, we depict the generic block diagram of the proposed scheme. There exist two sources of encrypted target pixels employed to embed data bits. Data bits at first are embedded in the encrypted white target pixels, $\mathbf{S} = \mathbf{P}_W^e$, and then in the encrypted black target pixels, $\mathbf{S} = \mathbf{P}_B^e$, as demonstrated in Subsection III-*B* by their own integration parameters $N_W$ and $N_B$. After embedding, the marked encrypted image, including the marked target pixels, $[\![\mathbf{P}_W^e]\!]$ and $[\![\mathbf{P}_B^e]\!]$, is created. Meanwhile, BRPs are just encrypted without any more modification. They will be employed to reconstruct other pixels at the recipient. They form 25% of the image.

TABLE I PERFORMANCE ANALYSIS OF THE PROPOSED SCHEME EMPLOYING 12 TEST IMAGES.

| Items | | F16 | Lena | Splash | House | Boat | Elaine | Lake | Peppers | Baboon | Stream | Aerial | APC |
|---|---|---|---|---|---|---|---|---|---|---|---|---|---|
| EC (bits) | | 161290 | 161290 | 161290 | 161290 | 86021 | 193548 | 96774 | 161290 | 86021 | 80645 | 86021 | 96774 |
| $N_W$ | | 1 | 1 | 1 | 1 | 2 | 1 | 2 | 1 | 2 | 2 | 2 | 2 |
| $N_B$ | | 2 | 2 | 2 | 2 | 3 | 1 | 2 | 2 | 3 | 4 | 3 | 2 |
| PSNR | | ∞ | ∞ | ∞ | ∞ | ∞ | ∞ | ∞ | ∞ | ∞ | ∞ | ∞ | ∞ |
| Retrieving $\mathbf{P}_W$ | HiR | 4 | 0 | 0 | 0 | 0 | 0 | 0 | 8 | 5 | 0 | 0 | 0 |
| | MeR | 136 | 80 | 4 | 163 | 44 | 27 | 2 | 161 | 1052 | 182 | 130 | 4 |
| Retrieving $\mathbf{P}_B$ | HiR | 0 | 0 | 0 | 0 | 0 | 1 | 0 | 0 | 13 | 0 | 0 | 0 |
| | MeR | 37 | 9 | 1 | 33 | 68 | 40 | 93 | 35 | 885 | 25 | 103 | 11 |

TABLE II SIX IMAGES OF BOWS2 DATABASE THAT HAVE THE MOST NUMBER OF HiR FOR EMBEDDING DATA IN $\mathbf{P}_W$.

| Items | | Images (.pgm) | | | | | |
|---|---|---|---|---|---|---|---|
| | | 6502 | 6501 | 6537 | 6498 | 6514 | 9373 |
| Retrieving $\mathbf{P}_W$ | HiR | 15 | 11 | 10 | 6 | 4 | 4 |
| | MeR | 947 | 923 | 778 | 943 | 415 | 460 |
| Retrieving $\mathbf{P}_B$ | HiR | 1 | 1 | 2 | 1 | 0 | 2 |
| | MeR | 285 | 259 | 413 | 415 | 60 | 170 |
| PSNR | | 52.39 | 49.38 | ∞ | 52.39 | 55.4 | ∞ |
| The number of deformed MSBs | | 6 | 12 | 0 | 6 | 3 | 0 |

Now, assuming an original image has the size of $\mathbb{P} \times \mathbb{Q}$, and regarding integration parameters of $\{N_W, N_B\}$, embedding capacity (EC) can be achieved by

$$EC = \mathbb{P} \times \mathbb{Q} \times \left(\frac{1}{2N_W} + \frac{1}{4N_B}\right) \quad (12)$$

Thus, by choosing $\{N_W = 1, N_B = 1\}$, the most possible EC that is $\frac{3}{4}(\mathbb{P} \times \mathbb{Q})$ would be achieved.

As shown in Fig. 8b, extracting data bits and reconstruction

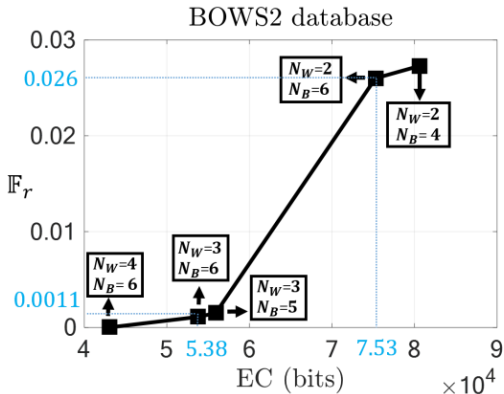

Fig. 9. Efficiency evaluation of the proposed scheme using 10000 test images of Bows2 database in different integration parameters, $\{N_W, N_B\}$. We just exploit $K_e$ to restore the original images. The symbol $\mathbb{F}_r$ is the failure rate.

of the original image are separately accomplished.

Data extraction is demonstrated in Subsection III-*C*. It is performed for the marked sources, $[\![\mathbf{P}_W^e]\!]$ and $[\![\mathbf{P}_B^e]\!]$.

Reconstructing the original image is initiated by decryption of the marked encrypted image using $K_e$. Therefore, BRPs, $\mathbf{P}_{\acute{B}}$, are completely recovered just by decryption. Having $\mathbf{P}_{\acute{B}}$, BTPs may be retrieved in a procedure described in Subsection III-*D*, where $[\![\mathbf{S}^d]\!] = [\![\mathbf{P}_B]\!]$. Having $\mathbf{P}_{\acute{B}}$ and $\mathbf{P}_B$, WTPs are recovered in similar procedure when $[\![\mathbf{S}^d]\!] = [\![\mathbf{P}_W]\!]$. BPP and WPP predictors are used to compute prediction-errors of $[\![\mathbf{P}_B]\!]$ and $[\![\mathbf{P}_W]\!]$, respectively. The risk of failure is included in the retrieval process as well.

If the encrypted target pixels are used sequentially for integration, the MSBs of pixels in rougher regions are considered as a set of integrated MSBs that rises possibility of failure in the reconstruction process. Thus, by scrambling the encrypted target pixels, we can mitigate the risk at the recipient.

IV. EXPERIMENTAL RESULT

The performance of the proposed separable VRAE algorithm is evaluated by conducting several experiments. Twelve grayscale images *F16, Lena, Splash, House, Boat, Elaine, Lake, Peppers, Baboon, Stream, Aerial* and *APC* from the USC-SIPI database are used as our test images. Also, BOWS2 original database, including 10000 greyscale images, are employed to confirm that the proposed algorithm can provide LR of the original image and error-free extraction of data bits. The size of all test images are 512×512. In experiments, the first and the last two rows and columns of the image are ignored in data embedding process. The peak signal-to-noise ratio (PSNR) is used to estimate the quality of the recovered image. PSNR= ∞ means LR of the original image.

Choosing integration parameters, $\{N_W, N_B\}$, is somewhat relevant to entropy of an image. The greater the entropy of the image, the greater value of $\{N_W, N_B\}$ must be selected for LR. However, the larger $\{N_W, N_B\}$ is chosen, the smaller EC would be achieved. In Table I, the precise $\{N_W, N_B\}$ for LR of the test images is tabulated. By taking $\{N_W = 1, N_B = 1\}$ for the *Elaine* image, the most possible EC in the proposed method that is 193548 bits, is achieved. On the other hand, the *Stream* image provides the lowest EC. In this image, the lowest possible integration parameters for LR are $\{N_W = 2, N_B = 4\}$. Also, in Table I, the risk of failure in reconstruction is evaluated by computing the number of subset that take either HiR or MeR. Although *F16*, *Peppers* and *Baboon* images are reconstructed perfectly, they include 4, 8 and 5 high risk subsets of $N_W$-pixels, respectively. Also, in reconstructing the subsets of $N_B$-pixels, *Baboon* and *Elaine* images take 13 and 1 subsets of high risk, respectively. Accounting the number of HiR and MeR, *Baboon* is more susceptive to failure in reconstruction; nonetheless, by choosing $\{N_W = 3, N_B = 5\}$ for data embedding in *Baboon*, no subset of HiR is left. It is obtained by paying the cost of reducing EC to 55913 bits.

In our scheme, we assume that data hider is completely blind to the original-content; thus, we cannot find out the least possible precise $\{N_W, N_B\}$ for LR. Nevertheless, in the next subsection, it is confirmed that a set of $\{N_W, N_B\}$ may be always found out for LR. It is worth mentioning that in the proposed scheme, error-free extraction of data bits is realized for all test images under any circumstances.

*A. Lossless retrieval*

In Fig. 9, we demonstrate the performance of the proposed scheme in LR and error-free extraction of data bits using 10000 test images of BOWS2 database. As shown, for various $\{N_W, N_B\}$, we accomplished the proposed algorithm to result in the number of failure in reconstruction. For example, in $\{N_W = 3, N_B = 6\}$, from 10000 test images, there exist 11 images that are not perfectly reconstructed and consequently the failure rate is $\mathbb{F}_r = 0.0011$. As can be seen, the more $\{N_W, N_B\}$ is preferred, the less EC and definitely the less failure rate would be achieved. In $\{N_W = 4, N_B = 6\}$, $\mathbb{F}_r$ is zero. Therefore, there exists permanently a set of $\{N_W, N_B\}$ to provide LR for all 10000 test images.

As shown in Fig. 9, modifying $N_W$ affects EC and $\mathbb{F}_r$ more than $N_B$. As an instance, in $N_B = 6$, increasing $N_W$ from two to three decreases EC and $\mathbb{F}_r$ as much as 21506 bits and 0.0249, respectively; while in $N_W = 3$, the increment of $N_B$ has no noticeable impact on EC and $\mathbb{F}_r$. Implementing the proposed algorithm for $\{N_W = 3, N_B = 6\}$, we sort all 10000 reconstructed images in a descending order by the number of their subsets which are HiR in retrieving $\mathbf{P}_W$. The first six sorted images are listed in Table II. In this table, PSNR demonstrates four images that are not reconstructed losslessly and thus, four out of all 11 failures, i.e. Fig. 9 for $\{N_W = 3, N_B = 6\}$, include in the first six sorted images. It confirms that the risk analysis is a proper assessment for evaluating LR, i.e. all 11 failures are included in the first 50 sorted images. In any failure, there exist some deformed MSBs that cannot be recovered correctly. The number of deformed MSBs are demonstrated in Table II for various images. The maximum one is just 12 bits for *6501.pgm* image; thus, there is not much bit error rate when LR does not realized.

*B. A visual demonstration of the proposed scheme*

Fig. 10 is a visual demonstration of the proposed scheme including the original, encrypted, marked encrypted and reconstructed images with their depicted histograms, respectively. We optionally employ AES in the counter mode, i.e. a stream cipher procedure, to encrypt the *Lena* test image. As shown in Fig. 10b, after encryption there exist no knowledge of the original content. Due to using a standard encryption algorithm like AES, as can be observed, the histogram of the encrypted image is uniformly distributed which guarantees the security of the proposed algorithm. Since in the proposed scheme, data hider is completely blind to the original-content, it absolutely preserves the content-owner privacy.

Fig. 10c is a description of the marked encrypted image. The histogram still has a uniform distribution. The original image is losslessly reconstructed as shown in Fig. 10d.

*C. Comparison with other schemes*

In Table III, the proposed scheme is compared with other separable VRAE schemes. After image encryption, the proposed scheme vacates room by "histogram modification of the integrated MSBs" (HMIMSB). Besides, schemes [27] and [29] vacate room by "compressing the least significant bits" (CLSB) and "encoding using LDPC code" (ELDPC), respectively. Meanwhile, they adopt some parameters to make a tradeoff between embedding capacity and lossless recovery. Their functionality is similar to the integration parameters of the proposed scheme. For a fair comparison between HMIMSB and those of CLSB and ELDPC, we employ these parameters in such a way that all test images are perfectly reconstructed. In this approach, we apply parameters $\{M = 4, S = 2, L = 271\}$, $\{q = 0.1\}$ for CLSB, ELDPC and $\{N_W = 2, N_B = 3\}$ for HMIMSB. Note that by choosing less $L, q$ and $\{N_W, N_B\}$, both EC and the risk of failure in reconstruction are increased, for all three schemes.

In ELDPC, for LR at the recipient, some information is needed that can be available just by having $K_d$. Therefore, as demonstrated in Table III, for this scheme just a high quality version of the original image can be retrieved without having $K_d$ while in the proposed scheme the original image can be reconstructed perfectly just by having $K_e$. Besides, HMIMSB can achieve more EC than ELDPC.

As described in Table III, in comparison to CLSB, the embedding capacity is significantly improved. Similar to ELDPC, their algorithm is dependent on having $K_d$ for LR of the original image.

In Wu's scheme [32], using data hider key, they select some target pixels in the encrypted image to embed data bits and employ altered CB predictor to reconstruct the original image. Their methods do not guarantee LR. In the experiment, their algorithm fails to perfectly reconstruct the original image for five test images out of nine ones. As illustrated in Table III, we improve their algorithm by designing a procedure to guarantee LR by employing histogram modification and MSBs integration. However, we pay the cost by less EC. All discussed schemes as well as the proposed scheme provide an error-free data bits extraction while the proposed scheme is the only one

TABLE III EFFICIENCY COMPARISON BETWEEN THE PROPOSED SCHEME AND OTHER SEPARABLE VRAE ONES FOR NINE TEST IMAGES.

| Schemes | Items | images | | | | | | | | |
|---|---|---|---|---|---|---|---|---|---|---|
| | | *F16* | *Lena* | *Splash* | *House* | *Boat* | *Elaine* | *Lake* | *Peppers* | *Baboon* |
| Zhang2012 [27] (CLSB) | EC (bits) | 1920 | 1920 | 1920 | 1920 | 1920 | 1920 | 1920 | 1920 | 1920 |
| | LR | Pass | Pass | Pass | Pass | Pass | Pass | Pass | Pass | Pass |
| | LR just by $K_e$ | Fail | Fail | Fail | Fail | Fail | Fail | Fail | Fail | Fail |
| **Wu [32]** | **EC (bits)** | **130050** | **130050** | **130050** | **130050** | **130050** | **130050** | **130050** | **130050** | **130050** |
| | **LR** | **Pass** | **Pass** | **Pass** | **Fail** | **Fail** | **Pass** | **Fail** | **Fail** | **Fail** |
| | **LR just by $K_e$** | **Pass** | **Pass** | **Pass** | **Fail** | **Fail** | **Pass** | **Fail** | **Fail** | **Fail** |
| Qian [29] (ELDPC) | EC (bits) | 77376 | 77376 | 77376 | 77376 | 77376 | 77376 | 77376 | 77376 | 77376 |
| | LR | Pass | Pass | Pass | Pass | Pass | Pass | Pass | Pass | Pass |
| | LR just by $K_e$ | Fail | Fail | Fail | Fail | Fail | Fail | Fail | Fail | Fail |
| **Proposed Scheme (HMIMSB)** | **EC (bits)** | **86021** | **86021** | **86021** | **86021** | **86021** | **86021** | **86021** | **86021** | **86021** |
| | **LR** | **Pass** | **Pass** | **Pass** | **Pass** | **Pass** | **Pass** | **Pass** | **Pass** | **Pass** |
| | **LR just by $K_e$** | **Pass** | **Pass** | **Pass** | **Pass** | **Pass** | **Pass** | **Pass** | **Pass** | **Pass** |

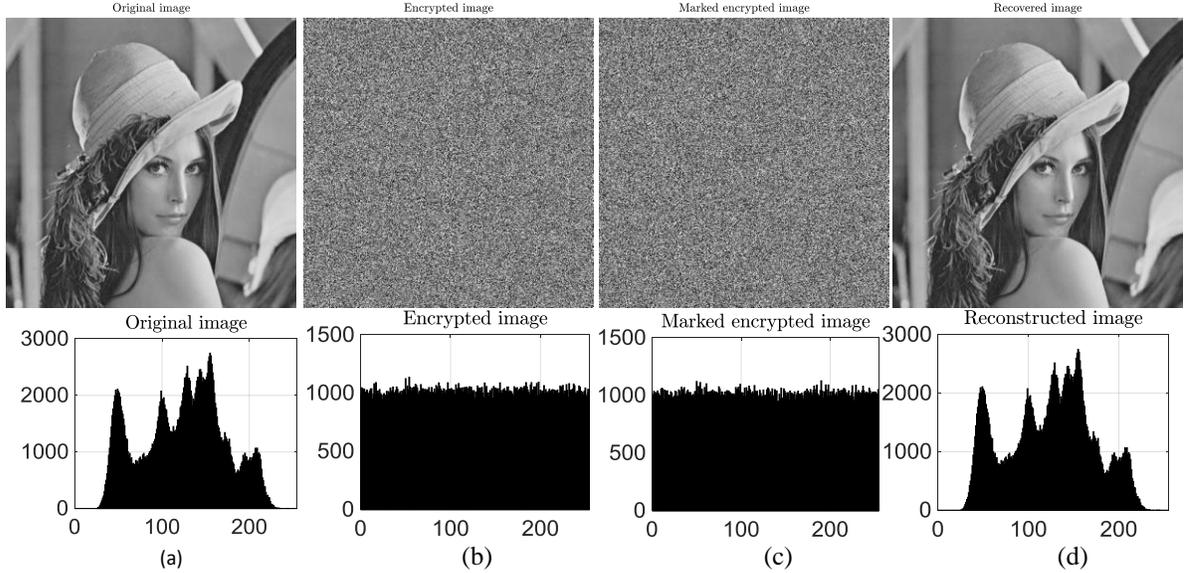

Fig. 10. A visual demonstration of the proposed scheme. (a) Original image. (b) Encrypted image. (c) Marked encrypted image, i.e. $EC = 161290$ bits. (d) Reconstructed image, $PSNR = \infty$. Histogram of the images is also described.

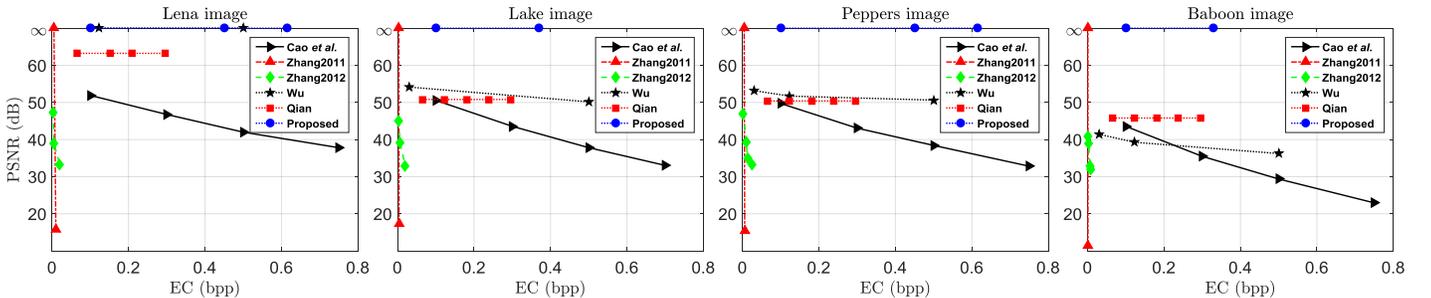

Fig. 11. PSNR comparison of the proposed scheme and other schemes such as Cao et al. [16], Zhang2011[30], Zhang2012 CLSB [27], Qian ELDPC [29], and Wu [32] for four test images.

that makes possible LR of the original image just by having the secret key ($K_e$).

In Fig. 11, the proposed scheme is compared to [16], [27], [29], [30] and [32] in terms of the quality of the reconstructed image. In this experiment, schemes [30] and [32] have both encryption and data hider keys while others just have encryption key. In Zhang's method [30], the reconstruction of the original image and extraction of data bits are joint while in our scheme it is separable. In this case, we improve not only EC but also the image quality. To embed data bits, Cao's scheme reserves room before encryption using patch-level sparse representation [16]. Generally, the embedding capacity of the proposed scheme is comparable with the Cao's method. A preprocessing is allowed before encryption in Cao's scheme while the proposed scheme is absolutely blind to the original-content. As shown in Fig. 11, ELDPC outperforms CLSB and the Cao's scheme in term of PSNR. From LR point of view, our proposed scheme outperforms ELDPC. In comparison with Wu's scheme, for *Peppers* and *Lena* images, we reach to higher EC. In *Lake*, *Peppers* and *Baboon* images, they achieve the PSNR of less than 55 dB even by using both keys while in our scheme LR is performed just by $K_e$.

## V. Conclusion

In this paper, by comparing different predictors, it is demonstrated that WPP is a better choice to reduce the probability of failure in reconstructing the original image. Moreover, BCP predictor is employed to increase the embedding capacity. By prediction-error analysis, we just choose the MSB of the encrypted target pixels for data embedding. These MSBs are integrated to be more robust against failure of reconstruction when they are modified to embed data bits. Employing histogram modification of the integrated MSBs, we vacate room for data embedding. At the recipient, data bits are faultlessly extracted and the original image is losslessly reconstructed by separate procedures. We employ the risk analysis evaluating LR. The proposed method improves other separable VRAE schemes thanks to using MSBs integration and histogram modification. As a future work, we are willing to improve the embedding capacity.